\documentclass[aps,prc,showpacs,nofootinbib,twocolumn]{revtex4}

\usepackage{epsfig}
\usepackage{graphicx,amsmath}
\newcommand\ba{\begin{eqnarray}}
\newcommand\ea{\end{eqnarray}}

\newcommand{\be}{\begin{equation}}
\newcommand{\ee}{\end{equation}}

\newcommand{\bas}{\begin{eqnarray*}}
\newcommand{\eas}{\end{eqnarray*}}
\begin{document}
\title{\bf \large Search for Low Mass Exotic leptonic or bosonic structures}

\author{
B. Tatischeff$^{1,2}$\thanks{e-mail : tati@ipno.in2p3.fr}\\
$^{1}$CNRS/IN2P3, Institut de Physique Nucl\'eaire, UMR 8608, Orsay, F-91405\\
$^{2}$Univ Paris-Sud, Orsay, F-91405, France\vspace{3.mm}\\
E. Tomasi-Gustafsson$^{3}$\thanks{e-mail : etomasi@cea.fr}\vspace*{1.mm}\\
$^{3}$DAPNIA/SPhN, CEA/Saclay\\ 91191 Gif-sur-Yvette Cedex, France}

\vspace*{1cm}
\begin{abstract}
Recently, several papers discussed the existence of a low mass leptonic structure. It was suggested that the $\Sigma^{+}$ disintegration: 
$\Sigma^{+}\to$pP$^{0}$, P$^{0}\to\mu^{-}\mu^{+}$ proceeds through an intermediate particle
P$^{0}$ having a mass close to M$\approx$~214.3~MeV.
The present work intends to look at other available data, in order to observe the eventual existence of a small peak or shoulder, at a mass close to M=214.3~MeV, which can strengthen the existence of a state produced by two leptons of opposite electric charge.  
\end{abstract}
\maketitle
\section{Introduction} 
The $\Sigma^{+}$ disintegration: $\Sigma^{+}\to$pP$^{0}$, P$^{0}\to\mu^{-}\mu^{+}$  was studied at Fermilab by H. Park {\it et al.}  \cite{park}. The data were taken by the HyperCP (E871) Collaboration. The authors observed a narrow range of dimuon masses, and supposed that the decay may proceed via a neutral intermediate state P$_{0}$, with a mass M=214.3~MeV $\pm$ 0.5~MeV. 

Several theoretical works were done assuming the existence of this new particle. Deshpande 
{\it et al.}  \cite{desh} assume a fundamental spin zero boson, which couple to quarks with flavor changing transition s$\to$d$\mu^{+}\mu^{-}$. They estimate the scalar and pseudoscalar coupling constants and evaluate several branching ratios. Xiao-Gang He {\it et al.}  \cite{xiao} re-examine the disintegration mode within the standard model, and find that, to be consistent with observations, the particles have small widths, short lived do not interact strongly. Geng and Hsiao \cite{geng} found that the P$^{0}$ cannot be scalar or vector but pseudoscalar, and determine that the decay width should be as small as $\approx$10$^{-7}$MeV. Gorbunov and Rubakov \cite{gorb} discuss a sgoldstino interpretation of this possible particle.

The experimental observation was based on three events. 
In order to eventually strengthen this result by a direct observation, we look at already existing data and try to observe a possible signature of a small peak, or a small shoulder, at a mass not far from the mass of P$^{0}$.
Such mass can be observed, either in the invariant mass of two muons, M$_{\mu\mu}$, or in missing masses of different reactions, studied  with incident leptons as  well as with incident hadrons. However the signal, if any, is expected to be small. If looked for in missing mass data, it will be superposed to a relatively large tail of one pion missing mass. Therefore the signal - if any -  can only be observed in precise data, with large statistics, good resolution and small binning. Moreover, the mass range studied must be small, in the 180$\le$M$\le$250~MeV range. Such data are scarce and concern reactions studied at rather low incident energies, with good resolution. When we found a hint for a small effect, we read out  and reanalyzed the data. Several of such structures were selected and are presented below.

The pp$\to$ppX reaction was studied at Saturne (SPES3 beam line), at T$_{p}$=1520, 1805, and 2100~MeV \cite{bt2}. The missing mass displays a broad structure, in the mass range
280$\le$M$\le$580~MeV, unstable for different kinematical conditions and slightly oscillating \cite{jy}, previously called the ABC effect; it was analyzed as being due to a superposition of four mesonic states: M=310~MeV, 350~MeV, 430~MeV, and 495~MeV \cite{jy}.

Since the widths of the missing mass peaks increase for increasing spectrometer angles, we keep only the three lowest angle spectra at T$_{p}$=1520~MeV, add them, and show the resulting spectra in 
Fig.~1(a). The high counting rate allows to extract a clear peak at M$_{X}$=216.5~MeV. 

The reaction pp$\to$pp$\pi^{0}\pi^{0}$ was studied close to threshold at Celsius \cite{bilger}. The missing mass data, after integration over two channels, are shown in Fig.~1(b). The two $\pi^{0}$ phase space starts at M$_{X}\approx$240~MeV.  The events in the range 170~$\le$M$_{X}\le$ 240~MeV, are mostly physical as the  background contribution is estimated to be lower than 10 events/channel.

These data are fitted with a $\pi^{0}$ peak and two peaks, having the same shape as the $\pi^{0}$ peak, at M$_{X}$=182 and 220~MeV.

The pp$\to$ppX reaction was also studied at J$\ddot{u}$lich COSY-TOF \cite{roder}.  Both protons in the final state were detected in order to study the $\eta$ production. The data were read and shown in Fig.~1(c). Since they are given in the original work as a function of the missing mass squared with constant binning ($\Delta$M$_{X}$=0.002~GeV$^{2}$), they are plotted as given up to M$_{X}$=210~MeV, (empty circles), and for larger missing mass they are integrated over two channels (full circles). The peak corresponding to $\pi^{0}$ missing mass is fitted by a gaussian, at M$_{X}$=135~MeV ($\sigma$=67~MeV), and the data at larger missing mass are fitted with a polynomial. Two structures can be extracted, the first one at M=197~MeV, not valid statistically, and the second at M=224~MeV. They are very narrow, therefore, if fitted by only one structure, which includes both narrow structures, they result in a broad gaussian centered at M=214~MeV (dashed curve in Fig~1(c))
\begin{figure}[!ht]
\begin{center}                                                          
\scalebox{.45}[.5]{
\includegraphics[bb=33 23 520 520,clip=]{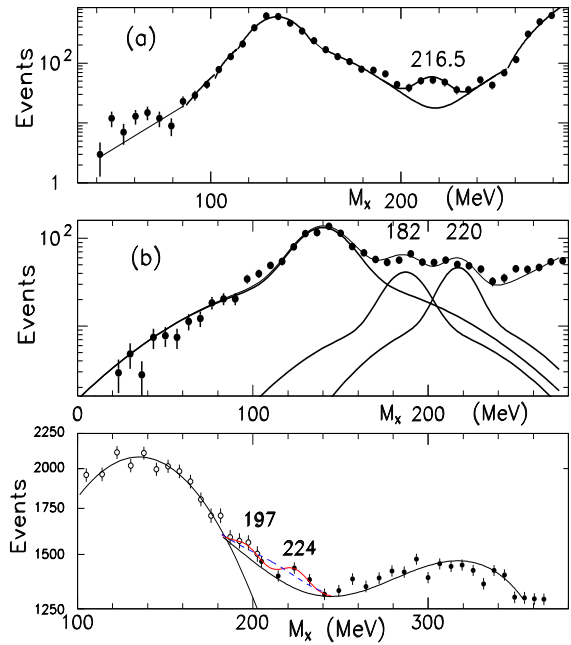}}
\caption{Insert (a): missing mass of the pp$\to$ppX reaction measured at Saturne (SPES3 beam line) at T$_{p}$=1520~MeV. Three spectra measured at $\theta_{pp}$=0$^{0}$, 2$^{0}$, and 5$^{0}$, are added. Insert (b): same reaction measured at Celsius \protect\cite{bilger}. Insert (c): Missing mass of the pp$\to$ppX reaction studied at J$\ddot{u}$lich COSY-TOF \protect\cite{roder}.}
\label{Fig1}
\end{center}
\end{figure}

The $\pi^{0}$ electroproduction at threshold  for Q$^{2}$=0.05~GeV$^{2}$ was measured at MAMI \cite{weis}. The missing mass spectrum, up to M$_{X}$=200~MeV is given in Fig.~3(b) of \cite{weis}, after background subtraction. The data are read, integrated by 4 channels and reported in Fig.~2(a). A peak at M$_{X}$=182~MeV is clearly observed. Indeed, the resolution in these data is as good as FWHM=2.2~MeV, as given in the $\pi^{0}$ peak (removed here to enhance the mass range discussed). The increase of the number of events between 48$\le$M$_{X} $~- M$_{\pi}\le$56~MeV is physical. The contribution from two pion production, is negligible at  M$\sim$ 180~MeV. 

The Roper resonance was studied at JLAB Hall A using  the p($\vec{e},e'\vec{p})\pi^{0}$ reaction \cite{sirca}. Two missing mass spectra were given at $\theta_{cm}$=90$^{0}$ and $\theta_{cm}$=-90$^{0}$. No shoulder is observed in this last spectra. The values of the spectra at at $\theta_{cm}$=90$^{0}$ are read and shifted in order to put the $\pi^{0}$ peak at his right mass, namely at M$_{X}$=135~MeV. Fig.~2(b) shows this spectrum fitted with a gaussian and two polynomials. A  small enhancement is observed at M$_{X}$=196~MeV.

The $\pi^{0}$ electroproduction on the proton was studied in Hall C at JLAB \cite{frolov}, in the region of the $\Delta$(1232) resonance via the  p(e,e'p)$\pi^{0}$ reaction. The authors give in Fig.~1 of Ref. \cite{frolov} an example of missing mass distribution for the reaction p(e,e'p)X. These data are read and reported in Fig.~2(c). The widths of all $\pi^{0}$ and $\eta$ peaks are related to their masses (proportional to 1/M). These widths fix the width of the small peak extracted at M=220~MeV. Several other peaks are introduced, following the results of the pp$\to$ppX reaction studied at SPES3 (Saturne) \cite{jy}. After introducing an arbitrary two-pion phase space, a contribution of the p(e,e'p)$\gamma$ reaction is observed around M$_{X}$=0.
\begin{figure}[!ht]
\begin{center}                                                          
\scalebox{.45}[.5]{
\includegraphics[bb=20 20 530 530,clip=]{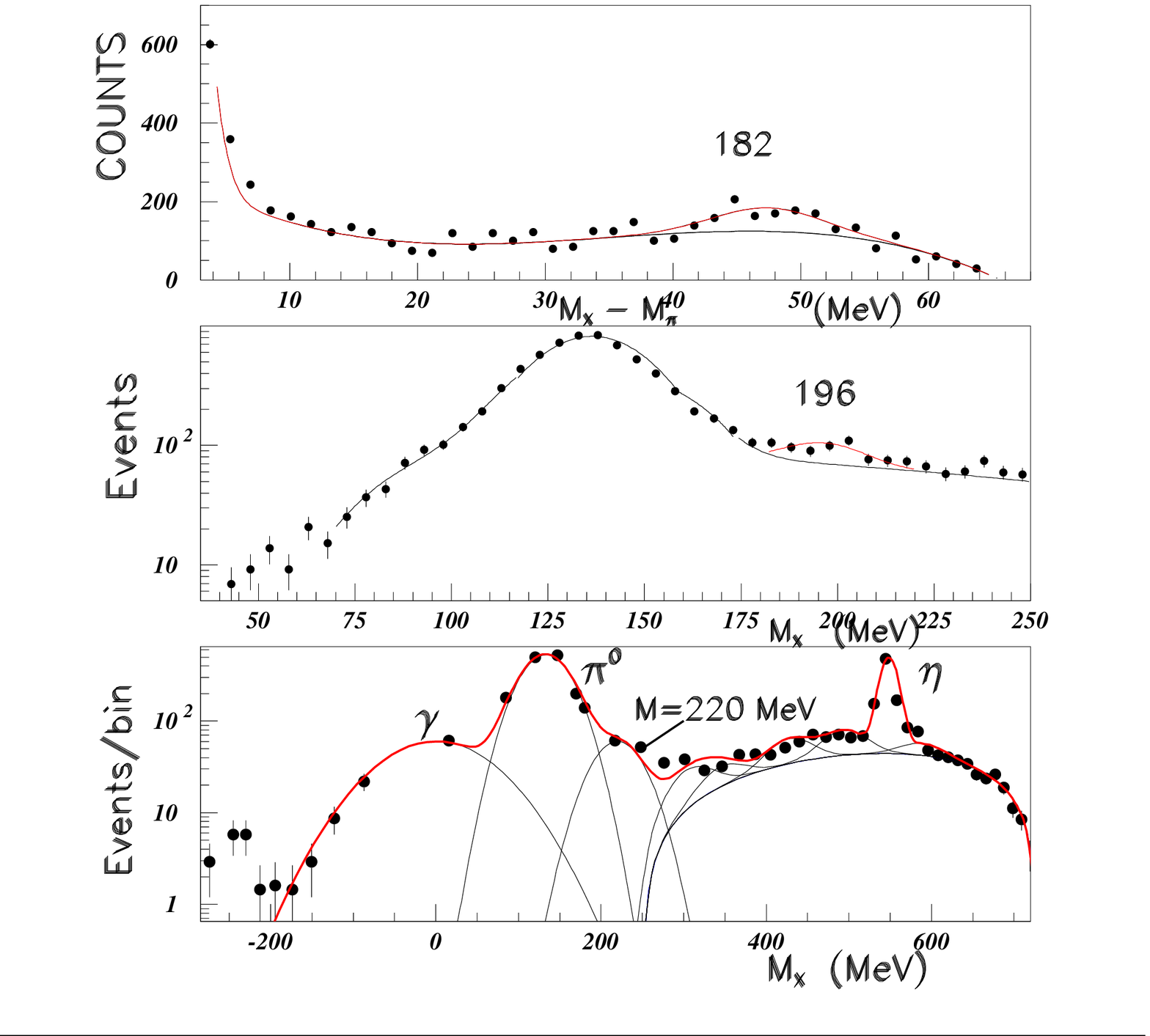}}
\caption{Insert (a): $\pi^{0}$ electroproduction at threshold, measured at MAMI \protect\cite{weis}. The missing mass spectra is integrated by 4 channels. 
 Insert (b): missing mass of the p($\vec{e},e'\vec{p})\pi^{0}$ reaction studied at JLAB Hall A at $\theta_{cm}$=90$^{0}$ \protect\cite{sirca}. 
Insert (c): missing mass of the p(e,e'p)X reaction measured at JLAB Hall C \protect\cite{frolov}.}
\label{Fig2}
\end{center}
\end{figure}
More detailed data from the same experiment \cite{frolovt}, are reported in  several spectra where structures can be extracted in the same missing mass range. The measurements were performed for two values of the four momentum transfer squared between the initial and the final electron, namely at Q$^{2}$=2.8~GeV$^{2}$ and Q$^{2}$=4~GeV$^{2}$.  The measurements were performed for a few values of P$^{0}_{p}$ and a few values of $\theta^{0}_{p}$. Four spectra are shown in Fig.~3 for  Q$^{2}$=4~GeV$^{2}$. 
Fig.~4 shows another selection of spectra corresponding to Q$^{2}$=2.8~GeV$^{2}$. In both figures empty circles correspond to Monte-Carlo simulations \cite{frolovt} and full circles correspond to data. In the missing mass range studied here an excess of counts can be seen between data and the simulation which were fitted by a polynomial. On the other the quantitative informations are given in Table~1. The discrepancy between data and simulation for M$_{X}\le$60~MeV  has been attributed to the Bethe-Heitler process (ep$\to$e'p'$\gamma$ reaction).
\begin{figure}[!h]
\begin{center}                                                          
\scalebox{.47}[.5]{
\includegraphics[bb=20 20 530 530,clip=]{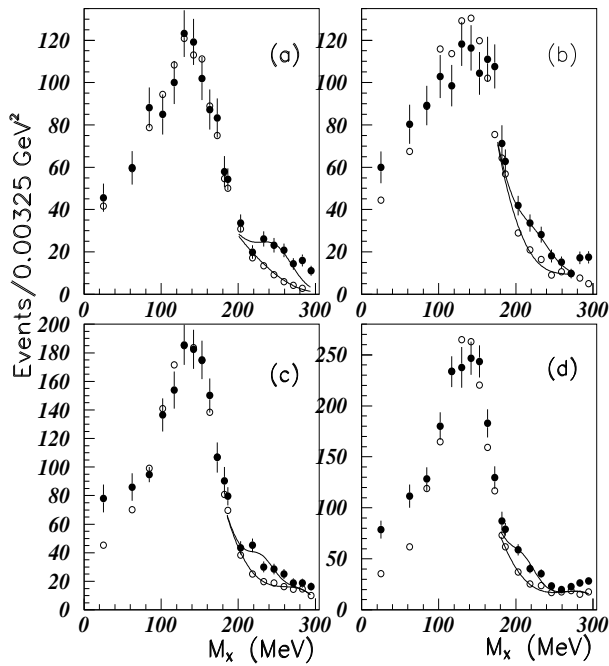}}
\caption{The missing mass of the  p(e,e'p)X reaction \protect\cite{frolovt} studied at JLAB Hall C at Q$^{2}$=4.0~GeV$^{2}$. The empty circles correspond to Monte-Carlo simulations; the full circles correspond to data. Inserts (a), (b), (c), and (d) correspond respectively to  p$^{0}_{p}$=2~GeV and $\theta^{0}_{p}$=23$^{0}$, 
 p$^{0}_{p}$=2~GeV and $\theta^{0}_{p}$=20$^{0}$, p$^{0}_{p}$=2.2~GeV and $\theta^{0}_{p}$=17$^{0}$, and 
 p$^{0}_{p}$=2.45~GeV and $\theta^{0}_{p}$=17$^{0}$.}
\label{Fig3}
\end{center}
\end{figure}
\begin{figure}[!ht]
\begin{center}                                                          
\scalebox{.47}[.5]{
\includegraphics[bb=20 20 530 530,clip=]{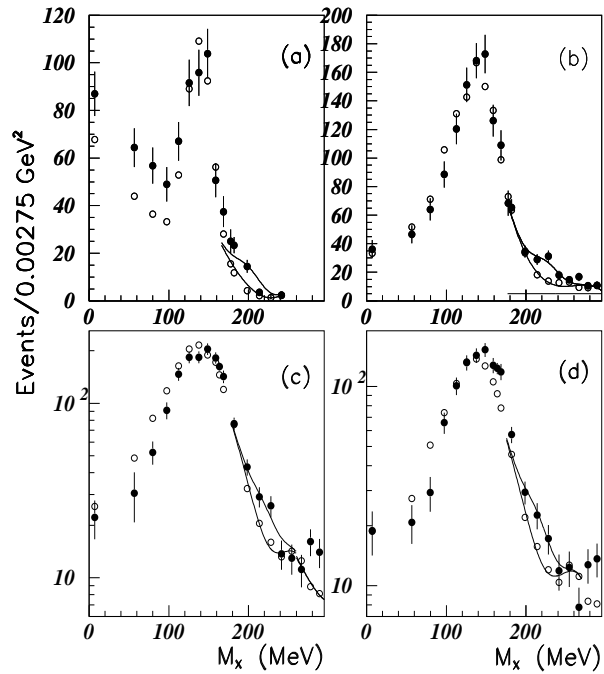}}
\caption{The missing mass of the  p(e,e'p)X reaction \protect\cite{frolovt} studied at JLAB Hall C at Q$^{2}$=2.8~GeV$^{2}$. The empty circles correspond to Monte-Carlo simulations; the full circles correspond to data. Inserts (a), (b), (c), and (d) correspond respectively to  p$^{0}_{p}$=1.9~GeV and $\theta^{0}_{p}$=33$^{0}$, 
 p$^{0}_{p}$=1.55~GeV and $\theta^{0}_{p}$=23$^{0}$, p$^{0}_{p}$=1.7~GeV and $\theta^{0}_{p}$=19$^{0}$, and 
 p$^{0}_{p}$=1.7~GeV and $\theta^{0}_{p}$=23$^{0}$.}
\label{Fig4}
\end{center}
\end{figure}
\vspace*{-3.mm}
%

We have looked at some existing data, in order to find evidence for the existence of a new boson, or a system of two leptons with opposite charge.
All spectra shown here, display a structure, but at slightly different masses. However, there is an indication of a possible regrouping around several mass values. The statistics is too low for giving an evidence if the results privilege one unstable mass or a few better defined masses.  We increase therefore the number of spectra studied, as those shown in Figs~3 and 4. The corresponding quantitative informations are summarized in Table~1. They favor a regrouping into several values; the same conclusion is favored by the existence of more than one peak in the same spectrum, as in Fig.~1(b). In summary these narrow structure masses (see Fig.~5), are tentatively observed at:\\
\hspace{3.mm}M=181$\pm$2~MeV (5 events),\\
\hspace{3.mm}M=198$\pm$2~MeV (5 events),\\
\hspace{3.mm}M=215$\pm$5~MeV (12 events),\\
\hspace{3.mm}M=227.5$\pm$2.5~MeV (5 events),\\
\hspace{3.mm}M=235$\pm$1~MeV (3 events).\\
\begin{figure}[!h]
\begin{center}                                                          
\scalebox{.47}[.5]{
\includegraphics[bb=47 284 520 520,clip=]{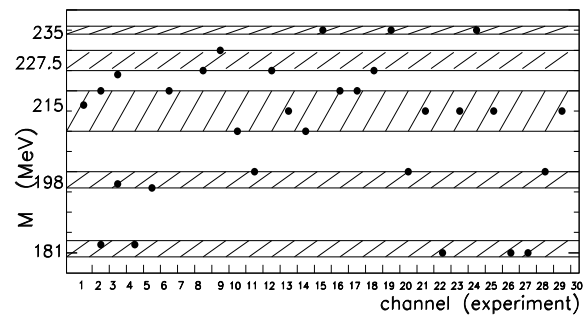}}
\caption{Masses of the weakly excited structures extracted from several experiments (see text and table~1).}
\label{Fig5}
\end{center}
\end{figure}
We notice that the range exhibiting the largest number of experimental mass structures, namely M=215~MeV, agrees with the value extracted at Fermilab: M=214.3~MeV \cite{park}.  There is also a qualitative evidence in favor of a structure at M$\approx$214~MeV. The pd$\to$pd$\eta$ reaction was studied at CELSIUS \cite{bilg}. Fig.~4 (lower frame) in Ref. \cite{bilg}  shows a scatter plot of M$_{\gamma\gamma}$ versus M$_{pd}$,
where a careful observation indicates a lightly dark range around M$_{\gamma\gamma}\approx$214~MeV.

The mass difference between the three lowest masses extracted before, is close to 
$\Delta$M=17.5~MeV. A mass gap  of 
\vspace*{-2.mm}
$$\Delta M=35~MeV=\displaystyle\frac{1}{2} \displaystyle\frac{m_e}{\alpha} $$
where $m_e$ is the electron mass and $\alpha$ is the fine structure constant,  was observed between several narrow hadronic exotic masses \cite{bt4}. Such gaps for leptons, mesons, and baryons were pointed out long time ago \cite{macgregor} and recently discussed in ref. \cite{palazzi}.

The mass difference observed here equals half this value. However, the masses of the narrow exotic mesons, baryons, and dibaryons were attributed to quark clusters \cite{bt2}, whereas the prevent peaks were first related to two muon states. In this case, the  narrow resonances observed here, would not be of the same nature.
\begin{table}[t]
\caption{Masses (in MeV) and number of standard deviations (S.D.) of the narrow peaks extracted around M$\approx$215~MeV, from the p(e,e'p)X reaction studied at JLAB Hall C \protect\cite{frolovt} for Q$^{2}$=4~GeV$^{2}$ (Fig.~3) and Q$^{2}$=2.8~GeV$^{2}$ (Fig.~4). The width of the peak is given by $\sigma$ (in MeV).  p$^{0}_{p}$ is in GeV, and $\theta$ is in degrees.}
\label{Table 1}
\vspace{5.mm}
\begin{tabular}{c c c c c c c c c c}
\hline
Fig.&insert&set in \protect\cite{frolovt}&Q$^{2}$&p$^{0}_{p}$&$\theta^{0}_{p}$&M&$\sigma$&S.D.\\
\hline
Fig.~3&(a)&10&4.0&2&23&250&22&9.5\\
          &(b)&11&4.0&2&20&225&22&4.9\\
          &(c)&21&4.0&2.2&17&230&17&5.2\\
          &(d)&5&4.0&2.45&17&210&17&6.3\\   
\hline
Fig.~4&(a)&14&2.8&1.9&33&200&17&1.9\\
            &(b)&19&2.8&1.55&23&225&17&4.0\\  
            &(c)&34&2.8&1.7&19&215&17&3.0\\          
            &(d)&36&2.8&1.7&23&210&17&2.9\\
\hline
\hline
&&12&4.0&2&17&235&22&5.1\\
&&14&4.0&1.8&17&220&22&2.6\\
&&15&4.0&1.8&20&220&17&4.9\\
&&16&4.0&1.8&23&225&17&1.5\\
&&19&4.0&2.2&23&235&17&5.3\\
&&20&4.0&2.2&20&200&17&5.9\\
&&4&4.0&2.45&14&215&17&3.5\\
&&6&4.0&2.45&20&180&17&3.7\\   
\hline
&&9&2.8&1.9&23&215&17&3.8\\
&&10&2.8&1.9&25&210&17&3.75\\
&&13&2.8&1.9&31&235&17&1.9\\
&&18&2.8&1.55&25&215&17&3.5\\
&&28&2.8&2.15&23&180&17&11\\
&&29&2.8&2.15&21&180&17&8.6\\
&&30&2.8&2.15&19&200&17&3.4\\
&&33&2.8&1.7&17&215&17&2.8\\            
\hline
\hline
\end{tabular}
\end{table}

The two lower masses M$_{\mu\mu}$ experimentally observed, are lower than the mass of two-muon disintegration threshold. A comparable situation was already observed in the baryon spectrum: two states, below the pion disintegration threshold, were extracted in the missing mass M$_{X}$ of the pp$\to$p$\pi^{+}$X reaction \cite{bt2}. The possibility to associate such gap with a boson of M=17.5~MeV, compares favorably to the model of Ref. \cite{walcher} which suggested the existence of a "genuine" Goldstone Boson with a mass close to 20~MeV.

Some results shown here are more qualitative, some results are statistically significant.  We have shown that the discussed structures can be observed in missing mass spectra from different reactions. All discussed data proceed from missing mass experiments, therefore the leptonic or bosonic nature of these structures can not be determined.

\end{document}